\documentclass[preprint,showpacs,preprintnumbers,amsmath,amssymb]{revtex4}

\usepackage{graphicx}
\usepackage{subfigure}
\usepackage{dcolumn}
\usepackage{bm}
\usepackage{epsfig}
\usepackage{color}
\usepackage{amsmath}
\usepackage{amsfonts}
\usepackage{amssymb}
\newcommand{\bra}[1]{\langle #1|}
\newcommand{\ket}[1]{|#1\rangle}
\newcommand{\braket}[2]{\langle #1|#2\rangle}

\begin{document}

\title{Generating entanglement between quantum dots with different resonant frequencies based on Dipole Induced Transparency}
\author {Deepak Sridharan and Edo Waks}
\affiliation{Department of Electrical and Computer Engineering,
IREAP and  JQI, University of Maryland, College Park, Maryland
20742, USA}

\begin{abstract}
 We describe a method for generating entanglement between two spatially separated dipoles coupled to optical micro-cavities.
  The protocol works even when the dipoles have different resonant frequencies and radiative lifetimes.
 This method is particularly important for solid-state emitters, such as quantum dots, which suffer from large inhomogeneous
 broadening. We show that high fidelities can be obtained over a large
 dipole detuning range without significant loss of efficiency.
We analyze the impact of higher order photon number states and
cavity resonance mismatch on the performance of the protocol.

\end{abstract}

\maketitle

\section{1. Introduction}
Generation of entanglement between qubits is an important
operation for a large variety of applications in quantum
information processing. Such states can be used to the realize
schemes such as transmission of secret messages via quantum key
distribution\cite{Ekert:1991,Jennewein:2000} and teleportation of
quantum
information\cite{Bennett:1993,Bouwmeester:1997,Buttler:1998,Mattle:1996}.
The exchange of entanglement between two distant parties is also
required for implementation of quantum repeaters
\cite{Briegel:1998} which use a combination of entanglement
swapping and entanglement purification\cite{Dur:1999} to achieve
unconditional secure communication over arbitrarily long
distances.

To date, a variety of methods have been proposed for creating
entanglement between spatially separated nodes.  One of the most
common methods is to transmit entangled photons generated by
parametric down-conversion\cite{Kwiat:1995}.  Entanglement
protocols for atomic systems have also been
proposed\cite{Jaksch:1999,Duan:2001,Childress:2006,Barrett:2005,Loock:2006}.
Atom entanglement has the advantages that quantum information can
be stored for long time periods, which is important for long
distance quantum networking.

Semiconductor based approaches to quantum information processing
are currently an area of great interest because they offer the
potential for a compact and scalable quantum information
architecture. Furthermore, solid-state emitters such as
semiconductor quantum dots (QDs),
 can be coupled to ultra-compact cavity waveguide systems to form highly integrated quantum systems
 \cite{Dirk:2007,Dirk:2005}. A major challenge in using solid-state emitters
  is that they suffer from
enormous inhomogeneous broadening, typically caused by emitter
size variation and strain fields in the host material. The
inhomogeneous broadening makes it difficult to find two emitters
with identical emission wavelengths. Protocols to date for
generating atom entanglement require the dipoles to emit
indistinguishable photons, and are thus difficult to implement in
semiconductor systems. In order to implement quantum networking in
semiconductors, we need a protocol that works even when the
dipoles emit photons that are distinguishable.

In this paper, we describe a protocol for creating entanglement
between two dipoles with different radiative properties, such as
different emission wavelengths or radiative lifetimes. The
proposed protocol uses Dipole Induced Transparency (DIT) to
achieve the desired entanglement which occurs when a dipole is
coupled to an optical cavity\cite{Waks:2006}.  When the coupling
is sufficiently strong, the dipole can switch a cavity from being
highly transmitting to highly reflecting.  The switching contrast
is determined by the atomic cooperativity, which is the ratio of
the lifetime of the uncoupled emitter  to the modified lifetime of
the cavity-coupled emitter. Enhancement of spontaneous emission
has been observed in semiconductor emitters coupled to a variety
of different micro-cavity
architectures\cite{Dirk:2005,Gerard:1998,Moreau:2001,Yoshie:2004,Badolato:2005,Reithmaier:2004}.
Modification of cavity reflectivity by coupling a quantum dot to a
photonic crystal nanocavity has been recently
observed\cite{Englund:2007,Kartik:2007}.

In section 2, we describe the basic protocol under idealized
assumptions that all fields can be expanded to first order in
photon number and that the two cavities have the same resonant
frequencies.  Section 3 then considers the effect of higher order
photon number states on the efficiency and fidelity of
entanglement.  The impact of non-linear behavior away from the
weak excitation limit is also discussed.  In section 4 we
investigate the effect of cavity frequency mismatch on the
entanglement.  Finally, in section 5 we perform a precise
numerical simulation of the entanglement generation protocol for
one specific implementation using the exciton bi-exciton cascade
of a single Indium Arsenide(InAs) quantum dot.  Numerical results
from recent experimental work is used to show that this may be a
promising method for achieving entanglement between two QDs for
the first time.

\section{2. Protocol for entanglement generation}

\begin{figure}[!h]
\centering
\includegraphics[width = 90mm,height = 60mm]{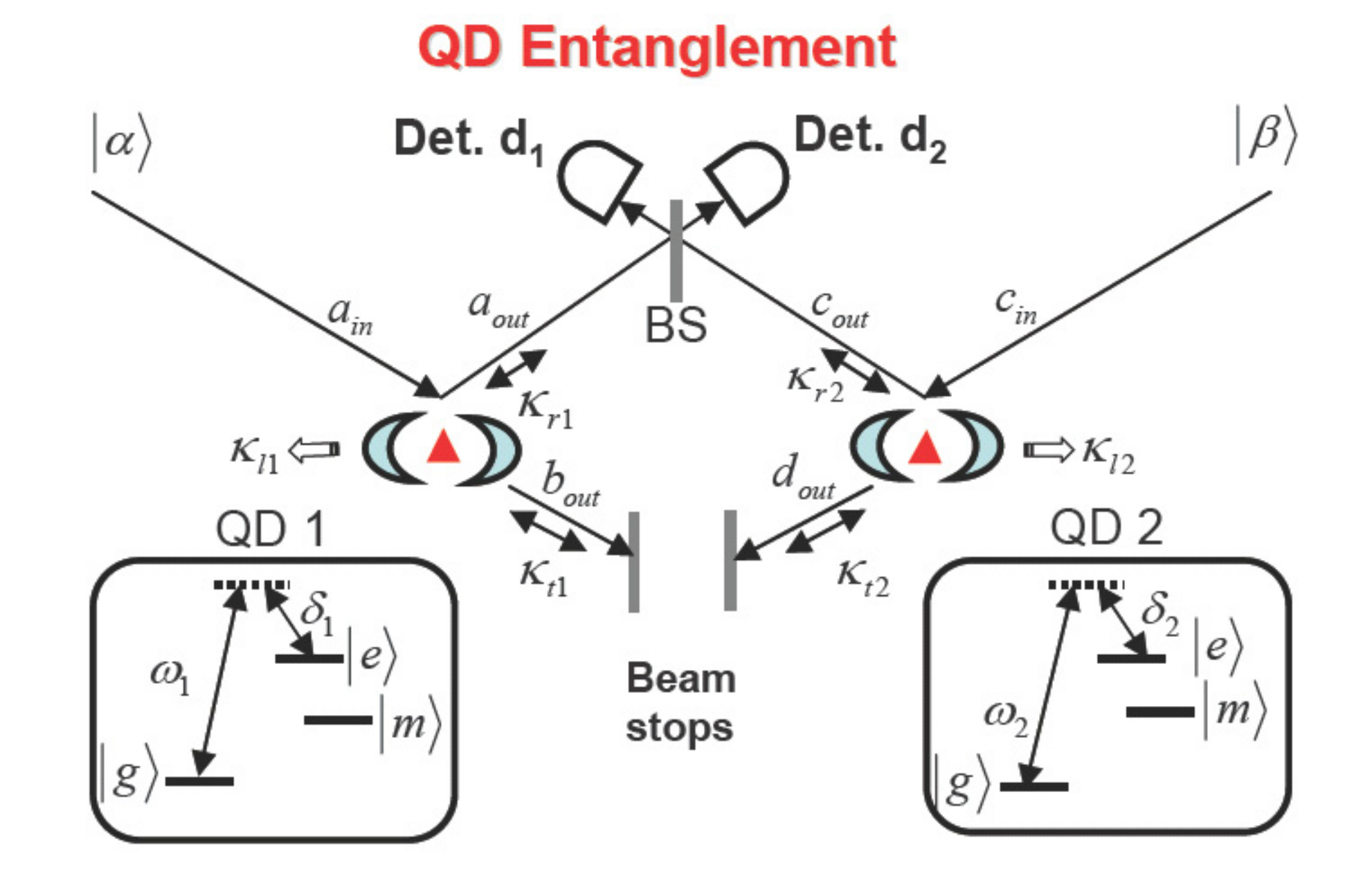}
\caption{Schematic of cavity waveguide system for generating
entanglement between two spatially separated dipoles using DIT }
\label{figure_1}
\end{figure}

The schematic for generating entanglement between two spatially
separated dipoles that emit distinguishable photons is shown in
Fig 1. Each qubit consists of a dipole coupled to a double sided
cavity. Each dipole is assumed
 to have three states: a ground state, a
long lived metastable state and an excited state, which we refer
to as $\ket{g}$, $\ket{m}$ and $\ket{e}$ respectively.  The states
$\ket{g}$ and $\ket{m}$ represent the two qubit states of the
dipole. The transition from the ground state to the excited state
for dipole 1, may be detuned by $\delta_1$ from the resonant
frequency $\omega_1$ of cavity 1. Similarly, the transition from
the ground state to the excited state for dipole 2 may be detuned
by $\delta_2$ from the resonant frequency $\omega_2$ of cavity 2.
We assume that  when the dipole is in state $\ket{g}$ it couples
to the cavity mode via an optical transition to state $\ket{e}$,
while when it is in state $\ket{m}$ it does not optically couple
to the cavity mode. State $\ket{m}$ may be decoupled from the
cavity due to either spectral detuning or selection rules.
Although state $\ket{m}$ is illustrated in the diagram as having
an energy level that is in between states $\ket{g}$ and $\ket{e}$,
this is not required. The only requirement is that when the dipole
is in state $\ket{m}$ it is decoupled from the cavity. This point
will be analyzed in more detail when we consider using the
exciton-biexciton transitions of an InAs QD to achieve
entanglement.
  The desired level structure described above can be realized in a variety of solid-state material systems.
In semiconductor quantum dots one can use the exciton and
biexciton transitions\cite{Santori:2001}, as well as the
spin-based bright and dark exciton states\cite{Stevenson:2006}. In
addition, three level structures can also be achieved using
quantum-dot molecules\cite{Stinaff:2006}, charged quantum
dots\cite{Gammon:2007} and impurity bound excitons\cite{Fu:2005}.
Similar qubits states could also be realized in other materials
such as diamond using neutral and negatively charged nitrogen
vacancy defects\cite{Santori:2006,Pieter:2006}.

The decay rates of the two dipoles is given by $\gamma_1$ and
$\gamma_2$ respectively. To characterize the interaction between
the dipoles and the cavity modes, we define the operators
$\hat{\mathbf{\sigma_{1-}}}$ and $\hat{\mathbf{\sigma_{2-}}}$.
They represent the dipole lowering operators for the dipoles in
cavities 1 and 2 respectively. It should be noted that
$\hat{\mathbf{\sigma_{1-}}}$ and $\hat{\mathbf{\sigma_{2-}}}$
represent the dipole lowering operators for the g-e transition.
Although state $\ket{m}$ is decoupled from the cavity, we still
define the dipole operators $\hat{\mathbf{\sigma_{m1-}}}$ and
$\hat{\mathbf{\sigma_{m2-}}}$ for the dipole in state $\ket{m}$.
These dipole are detuned by $\delta_{m1}$ and $\delta_{m2}$ from
their respective cavities.

 We define $\hat{\textbf{a}}_{in}$
and $\hat{\textbf{c}}_{in}$ as the two input modes,
$\hat{\textbf{a}}_{out}$  and $\hat{\textbf{c}}_{out}$ as the
reflected modes, and
 $\hat{\textbf{b}}_{out}$  and $\hat{\textbf{d}}_{out}$  as the transmitted
 modes to the two cavities, as illustrated in Fig. 1. The energy decay rate of
  cavity 1 into the reflected and transmitted modes is given by
$\kappa_{r1}$ and $\kappa_{t1}$ respectively. Similarly, the
energy decay rates of cavity 2 into the reflected and transmitted
modes is given by $\kappa_{r2}$ and $\kappa_{t2}$ respectively.
There is also the decay rate $\kappa_{l1}$ and $\kappa_{l2}$ into
the parasitic leaky modes that is due to losses such as material
absorption and out of plane scattering. The field inside the
cavities are represented by the cavity field operators
$\hat{\textbf{f}}_1$ and $\hat{\textbf{f}}_2$.

 The protocol works as follows. Both the dipoles are initialized to be in
 an equal superposition of qubit states $\ket{g}$ and $\ket{m}$. This can be achieved by first driving the dipoles into the
  lowest energy state by either waiting several radiative lifetimes or optical pumping.
   The qubit state can then be rotated by either a direct $\pi/2$ transition, or a raman transition\cite{Stievater:2001}.
    The choice of method depends on the specifics of the dipole and material system.
    Once the initialization step is complete, the initialized state of the two dipole system is given
     by $1/2(\ket{gg}+\ket{mm}+\ket{gm}+\ket{mg})$.

After the initialization of the dipoles, a weak coherent field
$\ket{\alpha}$ with frequency $\omega$ is inserted at input
$\hat{\textbf{a}}_{in}$. Simultaneously, another weak coherent
field $\ket{\beta}$ that is phase coherent with $\ket{\alpha}$
(i.e. originates from a common laser source) is injected at
$\hat{\textbf{c}}_{in}$. These input fields interact with the
cavity-dipole system. The interaction between the input field
$\hat{\textbf{a}}_{in}$ and cavity-dipole system 1 can be
characterized by the Heisenbergs equations of motion for the
cavity field operator $\hat{\textbf{f}}_1$ and the dipole lowering
operator $\hat{\mathbf{\sigma_{1-}}}$

\begin{equation}
\begin{split}
&\frac{d \hat{\textbf{f}_1}}{dt} = -(i\omega_0
+ (\kappa_{r1}+\kappa_{t1}+\kappa_{l1})/2)\hat{\mathbf{f}_1}-\sqrt{\kappa_{r1}}\hat{\mathbf{a_{in}}}-ig\mathbf{\sigma_{1-}}-ig\mathbf{\sigma_{m1-}} \\
&\frac{d\hat{\mathbf{\sigma_{1-}}}}{dt}=(-i(\omega_0+\delta_1)+\gamma_1)\mathbf{\sigma_{1-}}+ig\sigma_{z1}\hat{\textbf{f}_1}\\
&\frac{d\hat{\mathbf{\sigma_{m1-}}}}{dt}=(-i(\omega_0+\delta_m1)+\gamma_1)\mathbf{\sigma_{m1-}}+ig\sigma_{zm1}\hat{\textbf{f}_1}\\
\end{split}
\end{equation}
Similar equations can also be written for the interaction of the
input field $\hat{\textbf{c}}_{in}$ with cavity-dipole system 2.

The interaction between the input fields and the cavity-dipole
systems results in part of the field being transmitted into the
modes  $\hat{\textbf{b}}^{\dagger}_{out}$ and
$\hat{\textbf{d}}^{\dagger}_{out}$, while the remainder is
reflected into the modes $\hat{\textbf{a}}^{\dagger}_{out}$ and
$\hat{\textbf{c}}^{\dagger}_{out}$, or absorbed by the QD. The
amount of light reflected and transmitted is given by the cavity
reflection and transmission coefficients. Our analysis works in
the weak excitation limit, where predominantly the quantum dots
are populated in the ground state. In this limit,
$\langle\sigma_{z1}(t)\rangle \approx -1$. This also implies that
the population inversion for the dipole in state $\ket{m}$ is
close to $0$ i.e. $\langle\sigma_{zm1}(t)\rangle \approx 0$ Using
this limit in Eq. 1, we can derive the reflection and transmission
coefficients to be\cite{Walls:1994}

\begin{equation}
\begin{split}
 &r_{1}(\omega) = \frac{(-i\Delta\omega_{1}  + \frac{g^2}{-i(\Delta\omega_{1} - \delta_{1}) +
 \gamma_{1}})+ (\kappa_{r1}-\kappa_{t1}-\kappa_{l1})/2}{(-i\Delta\omega_{1}  + {(\kappa_{r1}+\kappa_{t1}+\kappa_{l1})/2}+ \frac{g^2}{-i(\Delta\omega_{1} - \delta_{1,2}) +
 \gamma_{1}})}\\
 &t_{1}(\omega) = \frac{\sqrt{\kappa_{r1}\kappa_{t1}}}{(-i\Delta\omega_{1}  +
 (\kappa_{r1}+\kappa_{t1}+\kappa_{l1})/2 + \frac{g^2}{-i(\Delta\omega_{1} - \delta_{1}) +
 \gamma_{1}})}
 \end{split}
 \end{equation}
 where
$\Delta\omega_1 = \omega - \omega_1$. These equations are obtained
for cavity-dipole system 1. The reflection and transmission
coefficients for dipole 2 are identical to dipole 1 in form, and
are obtained by substituting the dipole 2 parameters into Eq. 2.

 To get a better feel for Eq. 2, it is helpful to first consider
the simplified case where $\Delta\omega_{1} = 0$. Assume first
that the dipole is in state $\ket{g}$ and that the g-e transition
is resonant with the cavity such that $\delta_{1} = 0$.  We see
that maximum reflection and minimum transmission occurs for the
case when $\kappa_{r1} = \kappa_{t1} + \kappa_{l1}$, called the
critical coupling condition.  This condition ensures that no light
is reflected from the cavity when the incident field is directly
on cavity resonance.  We represent the decay rate $\kappa_{r1}$ at
critical coupling as $\kappa_1$. Hence, the transmission and
reflection coefficients simplify to t = 1/(1+C) and r = C/(1+C),
where C=$g^2/\gamma_1\kappa_1$ is called the atomic cooperativity.
If C$>>1$, which is the desired operation regime, then r=1 and all
of the light is reflected.  Now suppose the dipole is instead in
state $\ket{m}$ which is detuned from the cavity by $\delta_m$. We
then have t = 1/(1+CL) and r = CL/(1+CL), where
L=$\gamma_1/(\gamma_1+i\delta_m)$ is a Lorentzian function. If we
assume that either state $\ket{m}$ is highly detuned from the
resonance of the g-e transition ($\delta_m>>g^2/\kappa_1$) or the
transition is very weak due to selection rules (C$\approx 0$),
then t=1 and now the light is completely transmitted.  Thus, by
changing the state of the dipole from $\ket{g}$ to $\ket{m}$ we
can completely change the reflectivity of the cavity.

In a realistic system we cannot assume that the two dipoles are
resonant with their respective cavities, since in general they
will have different resonant frequencies.  Nor can we assume the
reflection and transmission coefficients will reach their ideal
limits because $\delta_m$ is not infinitely large and we usually
don't have perfect selection rules to cancel out the m-e
transition.  In this case we define for dipole 1, t$_{1}^g$ and
r$_{1}^g$ as the transmission and reflection coefficients when the
dipole is in state $\ket{g}$, and t$_{1}^m$ and r$_{1}^m$ when the
dipole is in state $\ket{m}$. We define r$_2^g$,t$_2^g$,r$_2^m$,
and t$_2^m$ analogously for dipole 2. These coefficients can be
calculated by plugging in the appropriate values corresponding to
the different transitions of the dipoles.  It is important to
emphasize that we do not assume that the $\delta$, g, $\gamma$,
and $\kappa$ are the same for both dipoles. The protocol we
describe works even if all of these parameters are different,
which is why it is so useful in semiconductors.

Before continuing, it is worth noting that in much of the
literature the atomic cooperativity C is often interchanged with
the Purcell factor, defined as the ratio of the lifetime of the
dipole inside the cavity to that of a dipole in bulk or free
space, which we denote $\gamma_{bulk}$.  Although the atomic
cooperativity is the correct parameter to use in the strictest
sense, it is a very difficult parameter to measure.  The Purcell
factor in contrast is easier to measure and almost always gives a
lower bound on the parameter C in any realistic system.  The
reason for this is that $\gamma_{bulk}$ is due both to radiative
and non-radiative decay.  In contrast the decay rate $\gamma$ is
the decay rate into non-cavity modes and is mainly due to
non-radiative processes, as radiation into modes other than the
cavity mode is highly suppressed.  Thus, outside of some atypical
cases where the cavity has two modes resonant with the QD, we
expect that $\gamma_{bulk}>\gamma$.  Therefore, in virtually all
cases C can be replaced by the Purcell factor to get a lower bound
on the performance of the system.

We first investigate the protocol under the assumptions that the
 resonant
frequencies of both the cavities are the
same($\omega_1=\omega_2$), and the input fields $\ket{\alpha}$ and
$\ket{\beta}$ are sufficiently weak that we may expand them to
first order in photon number. The initial state of the
system(dipoles and fields) is given by
$\ket{\Psi_i}=1/2(\ket{gg}+\ket{mm}+\ket{gm}+\ket{mg})
(\alpha\hat{\textbf{a}}^{\dagger}_{in}+\beta\hat{\textbf{c}}^{\dagger}_{in})$.
The fields, after interacting with the cavities, are transformed
according to cavity reflection and transmission coefficients. That
is, if dipole 1 is in state $\ket{g}$ then $\hat{\textbf{a}}_{in}
\rightarrow r_1^g \hat{\textbf{a}}_{out} + t_1^g
\hat{\textbf{b}}_{out}$, and if it is in state $\ket{m}$ then
$\hat{\textbf{a}}_{in} \rightarrow r_1^m \hat{\textbf{a}}_{out} +
t_1^m \hat{\textbf{b}}_{out}$. The transformation for photon in
$\hat{\textbf{c}}_{in}$ and dipole 2 is defined in a completely
analogous way. The reflected fields from the two cavities are
mixed on a 50/50 beamsplitter that applies the transformation:
$\hat{\textbf{a}}^{\dagger}_{out}\rightarrow(\hat{\textbf{d}}_{1}+\hat{\textbf{d}}_{2})/\sqrt{2},
\hat{\textbf{c}}^{\dagger}_{out}\rightarrow(\hat{\textbf{d}}_{1}-\hat{\textbf{d}}_{2})/\sqrt{2}$.
The final state of the QDs can be obtained by applying the cavity
and beamsplitter transformations.  If a detection event is
observed in detector $\hat{\textbf{d$_2$}}$, then the state of the
two QDs collapses to

\begin{equation}
\begin{split}
&\ket{\Psi_f}= \frac{1}{N}[(\alpha r_1^g-\beta
r_2^g)\ket{gg}+\alpha r_1^g\ket{gm} -\beta r_2^g\ket{mg}]
\end{split}
\end{equation}\
where $N^2=|\alpha r_1^g-\beta r_2^g|^2+|\alpha r_1^g|^2+|\beta
r_2^g|^2$.

In general, r$_1^g \ne$ r$_2^g$ because the dipoles have different
resonant frequencies.  However, we can correct for this mismatch
by properly adjusting the amplitudes of the fields. If the
amplitudes are selected such that
\begin{equation}
\alpha r_1^g=\beta r_2^g \label{match}
\end{equation}
 the state of the qubits
is projected onto $\ket{\Psi_-}=(\ket{gm}-\ket{mg})/\sqrt{2}$
which is an ideal entangled state. Thus, by properly choosing the
amplitude and phase of the input coherent fields $\ket{\alpha}$
and $\ket{\beta}$, we ensure that a detection at
$\hat{\textbf{d$_2$}}$ creates an entangled state of dipoles. Note
that the entanglement generation is accomplished despite the fact
that the two dipoles may have completely different resonant
frequencies or decay rates.

\section{3. Higher order photon numbers}

The matching condition $\alpha r_1^g=\beta r_2^g$ as given in Eq
\ref{match} gives a relationship between $\alpha$ and $\beta$, but
does not tell us how large to make $\alpha$. In general, we want
to make $|\alpha|^2$ as large as possible to improve the chances
of a detection event at $\hat{\textbf{d$_2$}}$.  The probability
of detecting a photon at detector $\hat{\textbf{d$_2$}}$ is
defined as the efficiency $\eta$ of the protocol. When the fields
$\ket{\alpha}$ and $\ket{\beta}$ are weak, efficiency of the
protocol is proportional to the intensity of the field at
$\hat{\textbf{d$_2$}}$ and can be derived to be $|\alpha
r_1^g|^2/4$. The factor $1/4$ appears because  $50\%$  of the
field is transmitted into the modes
$\hat{\textbf{b}}^{\dagger}_{out}$ and
$\hat{\textbf{d}}^{\dagger}_{out}$ and another $50\%$ is lost when
the beamsplitter splits the photons equally between
$\hat{\textbf{d$_1$}}$ and $\hat{\textbf{d$_2$}}$. We see that we
can achieve higher efficiencies by increasing input photon flux
rate $|\alpha|^2$. However, if we make $\alpha$ too large we can
no longer expand the fields to first order in photon number and
higher order photon number contributions will become important.

Higher order photon number contributions are undesirable because
they serve as a decoherence mechanism. In the ideal case where
only one photon is injected into the system, a detection event at
$\hat{\textbf{d$_2$}}$ ensures that there are no other photons in
the system which may carry "which path" information about the
state of the dipole. Now suppose we consider the second order
process of simultaneously injecting two photons into the input
ports $\ket{\alpha}$ and $\ket{\beta}$. In the ideal case (both
dipoles are on resonance with the cavities), if the state of the
two dipoles is $\ket{gm}$, cavity 1 will reflect its incident
photon while cavity 2 will transmit the second photon.  The
transmitted photon in cavity 2 will always keep track of the fact
that dipole 2 was in state $\ket{m}$, and this information cannot
be erased by the beamsplitter.  Thus, we expect the state to be
completely decohered when this happens.

We will now consider not only this specific case, but full
expansion of the coherent fields $\alpha$ and $\beta$ to all
photon numbers to see how the final state of the system is
affected. The initial state of the system is given by
$\ket{\Psi_i}=1/2(\ket{gg}+\ket{mm}+\ket{gm}+\ket{mg}\ket{\alpha}\ket{\beta}$.
The coherent states $\ket{\alpha}$ and $\ket{\beta}$ can also be
written as $\ket{\alpha}= D_1(\alpha)\ket{0}$ and $\ket{\beta}=
D_2(\beta)\ket{0}$. $D_1$ and $D_2$ are the displacement operators
and are given by
\begin{equation}
\begin{split}
&D_1(\alpha)=e^{\alpha \hat{\textbf{a}}^{\dagger}_{in} - \alpha^*
\hat{\textbf{a}}_{in}}\\
 &D_2(\beta)=e^{\beta
\hat{\textbf{c}}^{\dagger}_{in} - \beta^* \hat{\textbf{c}}_{in}}\\
\end{split}
\end{equation}
 The displacement operator provides as convenient way of writing the coherent states and includes all the higher order photon numbers
 contributions.

 The final state of system $\ket{\Psi_f}$ can be obtained by applying the cavity and beamsplitter transformations to the initial state
   $\ket{\Psi_i}$. After applying the transformations, the final state of the QDs is obtained by tracing out over the photon fields
   conditioned on a detection event at detector $\hat{\textbf{d$_2$}}$.  The state of the dipoles is therefore given by the reduced density matrix
\begin{equation}
\rho_{dipoles}=\frac{tr_{\mathrm{(fields)}}\{\braket{M}{\Psi_f}\braket{\Psi_f}{M}\}}{tr_{\mathrm{(dipoles
\& fields)}}\{\braket{M}{\Psi_f}\braket{\Psi_f}{M}\}}
\end{equation}
  The matrix $M = \sum_{n=1}^\infty\ket{n}_{d_2}\bra{n}$ is a
  positive
projector that projects the state of the system onto a subspace
containing at least one photon in $\hat{\textbf{d}}_{2}$. This
projection models the measurement performed by the photon counter,
which registers a detection event as long as there is at least one
photon in the detection mode.

Since the final state of QDs is mixed, we need a figure of merit
to measure how well the QDs are entangled. In this paper we use
the fidelity, which is  defined as the overlap integral between
the desired final state and the actual final state of the system.
In our protocol, the desired final state is the maximally
entangled Bell state $\ket{\Psi_-}$. Thus, the expression for
fidelity is $\bra{\Psi_-}{\rho_{dipoles}}\ket{\Psi_-}$. If the
actual final state is same as the desired final state, we have a
perfect entangled state and a fidelity of 1. A fidelity of 0.5
implies that the  state of the QDs is a random mixture of
$\ket{gm}$ and $\ket{mg}$ and completely decohered. An analytical
expression for the fidelity can be calculated by evaluating
$\rho_{dipoles}$ and averaging over the state $\ket{\Psi_-}$.  We
have carried out this calculation, but the expression for the
fidelity is messy and the math is involved.  The procedure for
calculating the fidelity along with the final analytical
expression are given in Appendix A.  The expression in the
appendix is used for subsequent calculations of fidelity.

 We also define the efficiency
$\eta$ as the probability of getting a detection at detector
$\hat{\textbf{d}}_{2}$. Mathematically this is given by the
expression, $\eta = {tr_{\mathrm{(dipoles \&
fields)}}\{\braket{M}{\Psi_f}\braket{\Psi_f}{M}\}}$. Using the
matching condition $\alpha r_1^g=\beta r_2^g$, the expression can
be simplified to $\eta = 0.5(1-e^{|\alpha r_1^g|^2/2})$.

\begin{figure}[h]
\centering
\includegraphics[width = 75mm,height = 40mm]{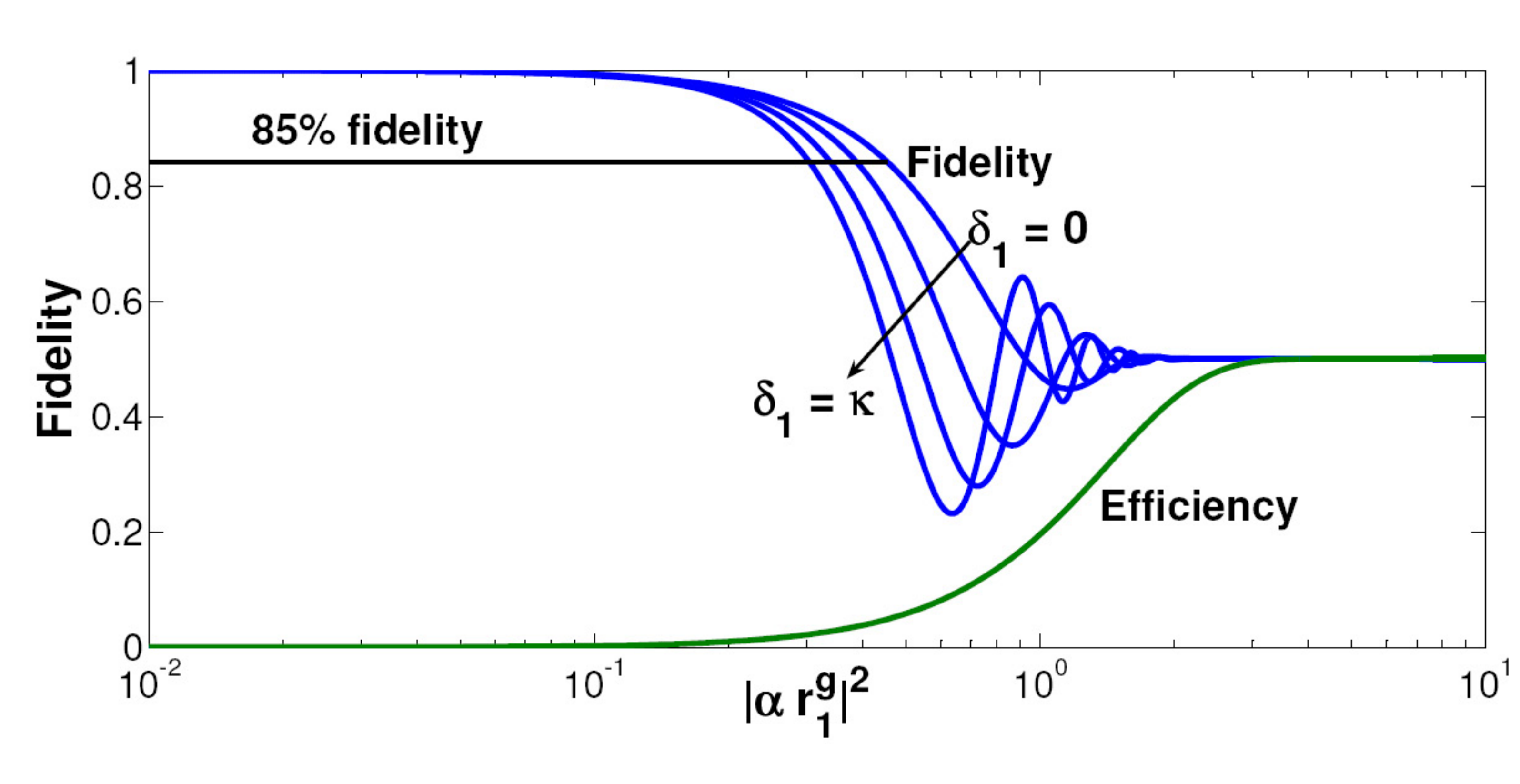}
\caption{Variation of fidelity and efficiency with $|\alpha
r_1^g|^2$ for different values of $\delta_1$.}\label{fig 3}
\end{figure}

 For the calculations in this
paper, we use parameters that are appropriate for InAs quantum
dots coupled to photonic crystal defect cavities. We represent the
total decay rate out of each cavity
$\kappa_{r1}+\kappa_{t1}+\kappa_{l1}$ and
$\kappa_{r2}+\kappa_{t2}+\kappa_{l2}$ as $\kappa$  and set it to
be equal to $100$ GHz. This corresponds to a cavity Q of 3300. We
set g = 20 GHz for both the quantum dots. We estimate dipole decay
rate $\gamma$ within the cavity to be 0.125 GHz, using reported
data that measured the lifetime of several quantum dots that were
placed inside a photonic crystal cavity, but heavily detuned
\cite{Englund:2005}. Using these values we calculate C to be 32
and the cavity-dipole systems to be $96.7\%$ reflective on
resonance. For the chosen values of g and $\kappa$, the
cavity-dipole systems are in the weak coupling
regime(g$<\kappa$/4). However, the analysis in this paper is
completely general and is equally valid also for the strong
coupling regime.  In Fig \ref{fig 3}, we plot both fidelity and
efficiency as a function of $|\alpha r_1^g|^2$. Fidelity is
plotted for four values of $\delta_1/\kappa$, ranging from 0 to 1,
with $\delta_2$ fixed at 0. Note that the efficiency is only a
function of $|\alpha r_1^g|$, so the plot of efficiency is the
same for all values of $\delta_1$. From Fig \ref{fig 3}, we see
that there is a tradeoff between fidelity and efficiency as we
increase $\alpha$. When $|\alpha r_1^g|^2<<1$ the fidelity is
close to 1, indicating an ideal entangled state, which is
consistent with our predictions in the weak field limit. In the
region 0.1$<|\alpha r_1^g|^2<$1, the fidelity quickly drops due to
the presence of higher photon number contributions.  In the limit
$|\alpha r_1^g|^2>>1$, the fidelity asymptotically approaches 0.5,
indicating the higher photon number contributions have completely
decohered the state.

When $|\alpha r_1^g|^2<<1$, the fidelity curves for different
values of $\delta_1$ nearly overlap. Fidelity stays close to 1 in
this region. However, in the region 0.1$<|\alpha|^2<$1, the
fidelity curves for different values of $\delta_1$ separate out.
There is a drop in fidelity with increase in dipole detuning from
$\delta_1=0$ to $\delta_1= \kappa$. Also, efficiency is a function
of $|\alpha r_1^g|$ alone
 and does not change with $\delta_1$. This implies that fidelity
decreases with increase in dipole detuning for a constant
efficiency.

In Fig \ref{fig 3}, we also plot a line of constant fidelity of
0.85.
 Note that for every value of $\delta_1$ there is a unique point on the
  plot corresponding to a fidelity of 0.85.  As $\delta_1$ increases, this point shifts to lower values of $|\alpha r_1^g|^2$.
  Since,
efficiency is a function of $|\alpha r_1^g|$, this in turn implies
a decrease in efficiency. Thus, it is important to consider how
the efficiency of the protocol changes for a fixed value of
fidelity.

\begin{figure}[!h]
\centering
\includegraphics[width = 75mm,height = 40mm]{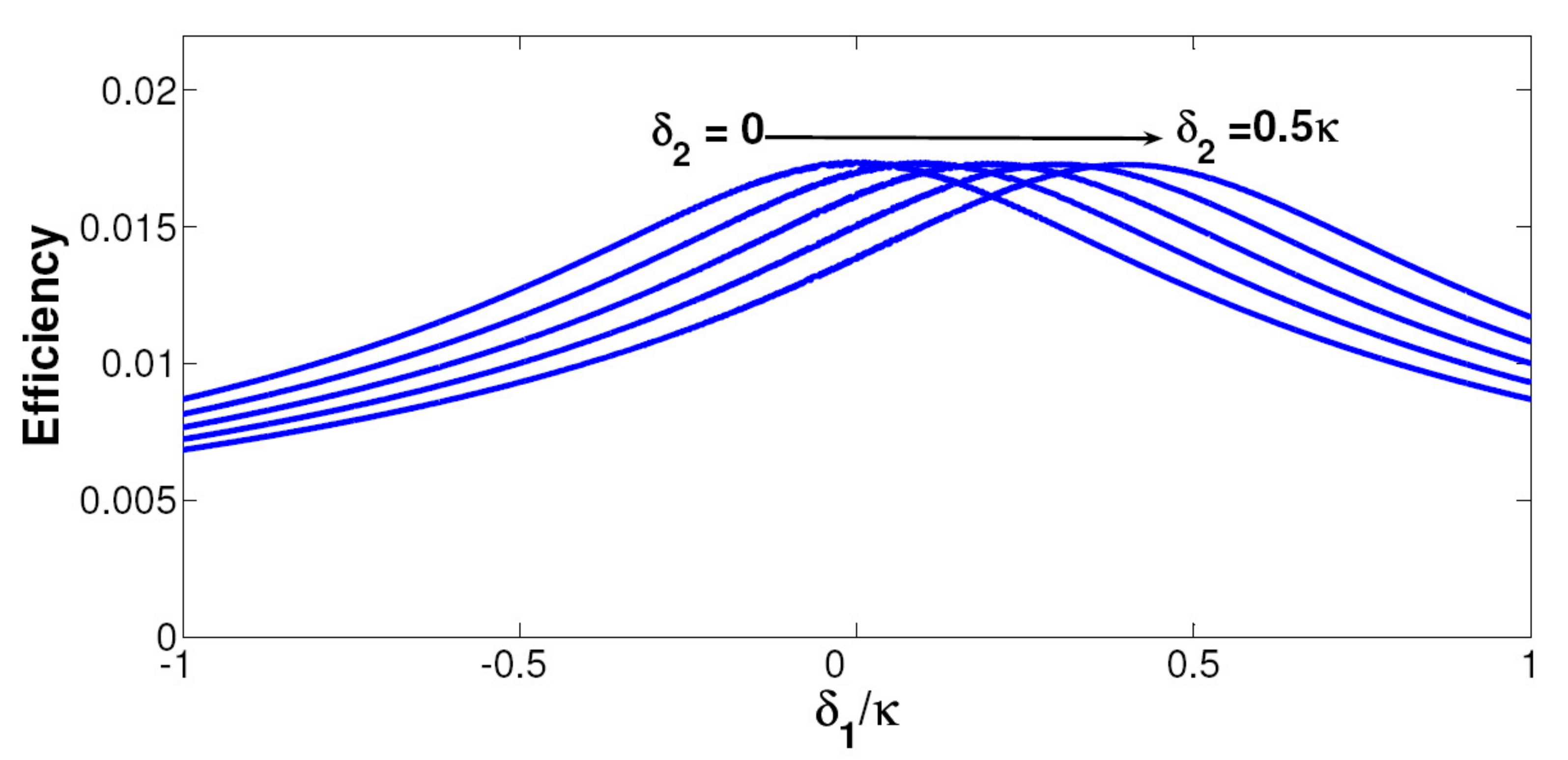}
\caption{Efficiency as a function of  $\delta_1/\kappa$ for
different $\delta_2$. Fidelity is fixed at 0.85}\label{fig 4}
\end{figure}

 To investigate this, we plot
efficiency as a function of $\delta_1/\kappa$ for several values
of $\delta_2$ for a constant fidelity of 0.85 in Fig \ref{fig 4}.
We see that even though there is a loss of efficiency, the change
is gradual and there is only a 50$\%$ reduction over a cavity
linewidth. Also, we would expect that if we added another detuning
$\delta_2$, efficiency would decrease. However, this does not
happen. From Fig \ref{fig 4} we see that the effect of $\delta_2$
is to shift the efficiency curves by the detuning $\delta_2$
without altering the shape. So, the protocol can be used to obtain
high efficiencies over a wide range of dipole detunings.

\section{ 4. Validity of Weak Excitation Limit}

In the protocol we describe, dipole detuning is compensated by
  adjusting the amplitude and phase of
the  input coherent fields until the matching condition  $\alpha
r_1^g = \beta r_2^g$ is satisfied. The more detuning we have, the
larger the amplitude required by the coherent field in order to
achieve the desired efficiency. It is possible that at some point,
the amplitudes required by the coherent fields will be so large
that the g-e transition of the QDs will be saturated leading  to
an optical nonlinearity and linewidth
broadening\cite{Garnier:2007}. Because of this, the cavity
reflection and transmission equations will depend on the pump
power and Eq 2 needs to be modified accordingly. However, our
protocol is intended to  work in the linear regime wherein the QDs
are unsaturated. This is possible only if the amplitude of the
input fields is within a certain limit called the weak excitation
limit. The weak excitation limit is defined as
$\langle\sigma_z(t)\rangle \approx -1$, which is equivalent to the
statement $\langle\sigma_+\sigma_-\rangle<<1$. and is necessary
for Eq. 2 to be valid. This condition puts a constraint on the
operation of the protocol.

In order to investigate the implication of the weak excitation
constraint, we start with the Heisenbergs equation of motion for
the cavity field operator $\hat{\textbf{f}_1}$ and the dipole
lowering operator $\hat{\mathbf{\sigma_1-}}$ given in Eq 1. We
will consider cavity-dipole system 1. Similar equations are also
applicable for cavity-dipole system 2.

Eliminating $\hat{\textbf{b}}$ from Eq 1, we have
\begin{equation}
[\frac{\kappa}{2}(i\delta_1+\gamma)-g^2]\hat{\mathbf{\sigma_-}}=-ig\sqrt{\kappa}\hat{\textbf{a}}^{\dagger}_{in}
\end{equation}
Using the fact the cooperativity index C is $g^2/\gamma \kappa>
1$, the equation can be further simplified  and multiplied with
its conjugate to obtain
\begin{equation}
\langle\sigma_+\sigma_-\rangle=\frac{g^2\kappa}{(g^4+\delta_1^2
\kappa^2/4)}
\langle{\hat{\mathbf{a}}^{\dagger}_{in}}\hat{\mathbf{a}}_{in}\rangle
\label{return}
\end{equation}

The parameter $\langle\sigma_+\sigma_-\rangle$ represents the
probability of the QD being in the excited state. In the weak
excitation limit, $\langle\sigma_+\sigma_-\rangle << 1$. We also
identify
$\langle{\hat{\mathbf{a}}^{\dagger}_{in}}\hat{\mathbf{a}}_{in}\rangle$
as the total flux of photons in the input field $\ket{\alpha}$.
Using this in Eq \ref{return}, the weak excitation constraint thus
puts a limit on $|\alpha|^2$ given by

\begin{equation}
\label{weaklimit} \frac{|\alpha|^2}{\tau_p} << \frac{g^2}{\kappa}
+ \frac{\kappa\delta_1^2}{g^2}
\end{equation} where $\tau_p$ is the pulse width of the laser.

 From Eq \ref{weaklimit}, we see that when there is no detuning $\delta_1$, the
flux of photons in the input field $\ket{\alpha}$ should be less
than the modified lifetime of the QD within the cavity
$\frac{g^2}{\kappa}$. This is understandable because, if the first
photon excites the QD and the second photon comes in before the QD
has decayed, we will no longer be in the weak excitation limit.
However, if the QD is off resonant from the cavity with detuning
$\delta_1$, not all the light that comes in couples to the QD.
Therefore, we will be able to pump the QDs with much more power
before we exceed the weak excitation limit. This is given by the
detuning dependent term $\frac{\kappa\delta_1^2}{g^2}$ in Eq
\ref{weaklimit}.

Eq \ref{weaklimit} conveys more than the weak excitation limit of
$\ket{\alpha}$.  If we apply the matching condition $\alpha r_1^g
= \beta r_2^g$  in Eq \ref{weaklimit}, we obtain a limit on the
flux of photons in the input field $\ket{\beta}$ given by

\begin{equation}
\label{weaklimitb} \frac{|\beta|^2}{\tau_p} << \frac{g^2}{\kappa}
+ \frac{\kappa\delta_2^2}{g^2}
\end{equation}
We recognize this as the weak excitation limit equation for the
field $\ket{\beta}$ which we would have obtained had we used the
Heisenbergs equations of motion for cavity-dipole system 2. This
implies that if we pick  $\ket{\alpha}$ such that it satisfies the
weak excitation limit of cavity-dipole system 1, the matching
condition automatically ensures that the flux of photons in
$\ket{\beta}$ is within the weak excitation limit of cavity-dipole
2.

Note that by making $\tau_p$ sufficiently long, we can always
ensure that the system is in the weak excitation limit and that
nonlinearities do no contribute.  However, because we are using
longer pulses the entanglement rate is reduced.  The rate of
entanglement generation is proportional to the rate at which the
cavity reflects photons, given by $ R = |\alpha r_1^g|^2/\tau_p$.
Using the upper limit on $|\alpha|^2/\tau_p$ from Eq
\ref{weaklimit} and cavity reflectivity $r_1^g$ from Eq 2, we get

\begin{equation}
\label{weakreflected} R << \frac{g^2}{\kappa}
\end{equation}
The above equation implies that the system will remain in the
linear weak excitation limit provided that the rate of reflected
photons is less the 1 photon per modified lifetime of the dipole.
Note that this result is true regardless of the detunings, and is
therefore valid in all cases.

It is instructive to compare the limits on the entanglement rate
imposed by nonlinearities to the limits imposed by which-path
information given in Section 3. The analysis of higher order
photon numbers in the previous section showed that reflected
photons $|\alpha r_1^g|<<1$ to have a high fidelity entangled
state between the QDs. In contrast, the analysis of weak
excitation limit in this section puts an upper bound on the rate
of the input photons in $\ket{\alpha}$ and $\ket{\beta}$ given by
$|\alpha r_1^g|^2/\tau_p << g^2/\kappa$ . Thus, the two analyses
are fundamentally different in that one limits the total number of
input photons and the other limits the rate of incoming photons.
Although one might expect the nonlinear limit analyzed in section
4 to be important, it turns out that the analysis of section 3 is
more restrictive, and is therefore the important limit to
consider.  To understand why, we first note that nonlinearities
can always be suppressed by increasing the pulse duration
$\tau_p$.  No mater how many photons we inject into the system, if
we make the pulses sufficiently long we will always be in the weak
excitation limit.  In contrast, which-path information does not
depend on pulse duration, and therefore cannot be suppressed.
Furthermore, in order to stay in the monochromatic limit (i.e. to
use the single frequency approximation) it has been shown in
previous work that the pulse duration must be longer than the
modified spontaneous emission lifetime of the
dipole\cite{Waks:2006}. If we combine this with the results of
section 3, which state that the number of reflected photons
$|\alpha r_1^g|^2<<1$, these two conditions already constrain us
to work in the regime where the $|\alpha r_1^g|^2/\tau_p <<
g^2/\kappa$.  Thus, we expect the entanglement to decohere due to
which-path information before the nonlinear behavior in section 4
is observed.  For this reason deviation from weak excitation does
not pose any additional restrictions to the protocol that were not
already present in the linear scattering regime.

\section{5. Effects of Cavity detuning}

In previous sections, we considered the idealized case where both
the cavities had identical resonant frequencies. However in
realistic systems, this will not be the case. Fabrication
imperfections may lead to slightly different resonances for the
two cavities.  Clearly, if even a small amount of mismatch between
the cavities were to result in no entanglement, the usefulness of
our protocol would be questionable.  Thus, it is important to
consider how sensitive the protocol is to cavity resonance
mismatch.

Now let's consider the case where the two cavities do not have the
same resonant frequency. The analysis of the protocol in the
presence of cavity detuning becomes  involved for two reasons.
First, it is no longer clear which frequency we should use for the
coherent fields $\ket{\alpha}$ and $\ket{\beta}$.  We do not know
whether to place it on resonance with one of the cavities or
somewhere in between. This can depend on both the cavity
separation $\Delta\omega_s$ and dipole detunings $\delta_1$ and
$\delta_2$.

Second, the matching condition used in the previous section
$\alpha r_1^g=\beta r_2^g$, is not guaranteed to be optimal. If a
detection event is observed in detector $\hat{\textbf{d$_2$}}$,
then the state of the two QDs is
\begin{equation}
\begin{split}
&\ket{\Psi_f}_{dipoles}= \frac{1}{N}[(\alpha r_1^g-\beta
r_2^g)\ket{gg}+(\alpha r_1^m-\beta r_2^m)\ket{mm}\\
&+\alpha r_1^m\ket{mg} -\beta r_2^m\ket{gm}]\\
\end{split}
\end{equation}\
where $N^2=|\alpha r_1^m-\beta r_2^m|^2+|\alpha r_1^m|^2+|\beta
r_2^m|^2$. The matching condition $\alpha r_1^g=\beta r_2^g$
ensures that we do not have any detection at
$\hat{\textbf{d$_2$}}$ if both the dipoles are in the state
$\ket{g}$. However,  the field amplitude at $\hat{\textbf{d$_2$}}$
if both the dipoles are in the state $\ket{m}$ i.e. ($\alpha
r_1^m-\beta r_2^m)$ is not compensated. This results in imperfect
destructive interference at detector $\hat{\textbf{d$_2$}}$. Thus,
there is a small probability of detection at
$\hat{\textbf{d$_2$}}$ when both the dipoles are in state
$\ket{m}$. This causes a loss of fidelity. In order to obtain the
state that comes closest to the desired final entangled state, we
must optimize the fidelity with respect to $\omega$, $\alpha$ and
$\beta$.

For calculating the effects of cavity detuning, we choose the
frequency midway between the two cavity frequencies as the
reference frequency $\Delta_{ref}$. Based on this reference
frequency, $\omega_1 = -\Delta\omega_s/2$ and $\omega_2 =
\Delta\omega_s/2$. Also, it will be easier if we define the dipole
detunings in terms of the reference frequency rather than the
cavity frequencies. We define $\Delta_1 = \delta_1 + \omega_1$ and
$\Delta_2 = \delta_2 + \omega_2$, which are the dipole detunings
of dipoles 1 and 2 with respect to the reference frequency located
midway between the two cavities. These definitions ensure that
when increasing the cavity separation $\Delta\omega_s$ we do not
affect the QDs. This is important because we can obtain
information about the effects of cavity detuning alone by making
these definitions.

Figure 4 plots the dependence of fidelity on the laser frequency
for several different values of $\Delta_1$. The cavity separation
$\Delta\omega_s = \omega_2-\omega_1$ is set to 50 GHz, and
$\Delta_2=0.25\kappa$. The figure is optimized over the real and
imaginary parts of $\frac{\alpha}{\beta}$. The value of the
maximum fidelity for the three curves occurs at three different
frequencies. The frequency at which we get maximum fidelity is the
optimal frequency $\omega$. The fidelity at that frequency is the
maximum fidelity that can be obtained for that particular
configuration of $\Delta\omega_s$, $\Delta_1$ and $\Delta_2$.

\begin{figure}[!h]
\centering
\includegraphics[width = 75mm,height = 40mm]{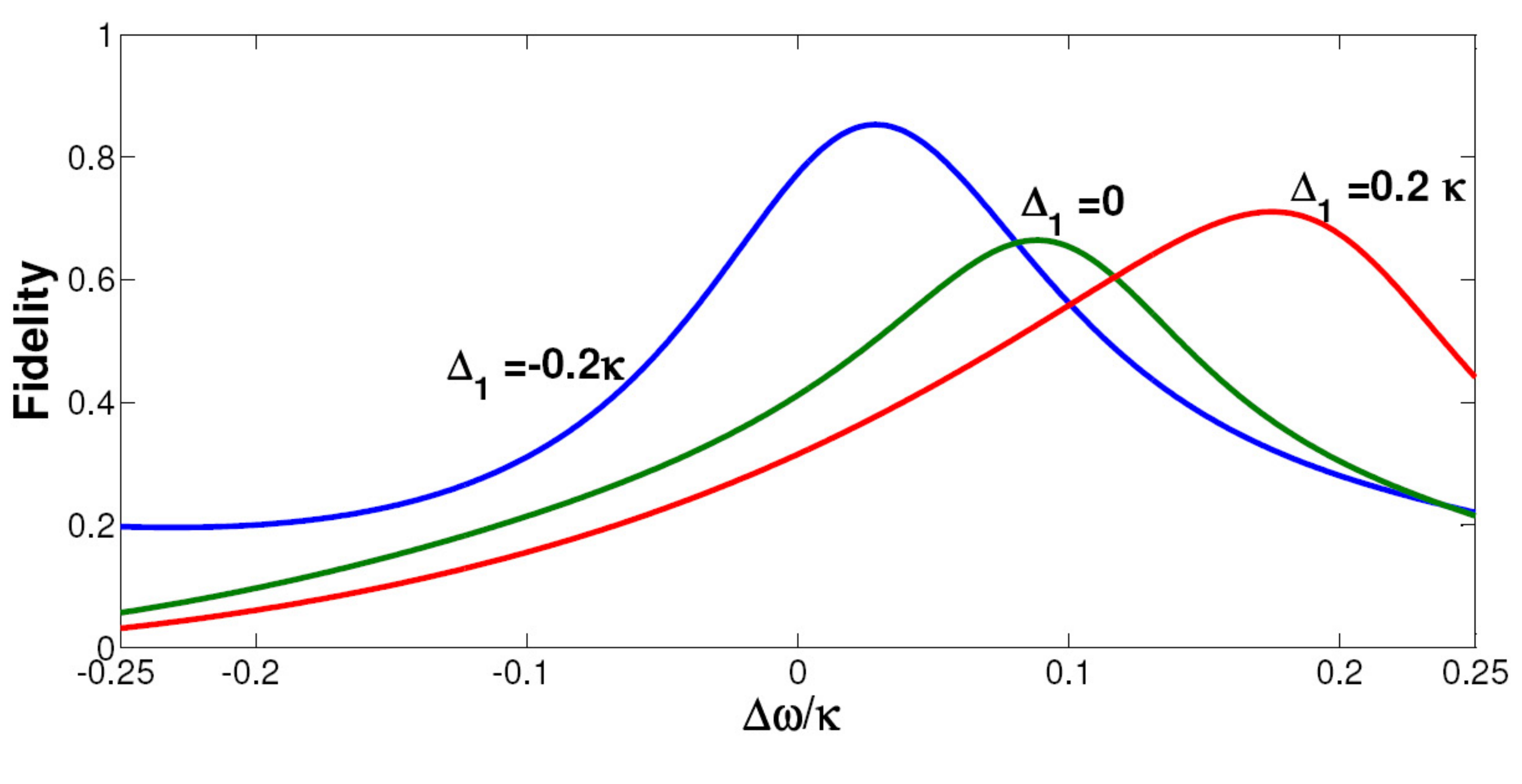}
\caption{a) Fidelity as a function of laser frequency for
different values of $\Delta_1$. Optimization is performed over the
real the imaginary parts of $\frac{\alpha}{\beta}$. Cavity
separation $\Delta\omega_s$ is set to 50 GHz and $\Delta_2 =
0.25\kappa$ GHz}\label{ FIG 5}
\end{figure}

In Fig. 5, we plot optimized fidelity as a function of cavity
detuning $\Delta\omega_s$ for different values of $\Delta_1$ with
$\Delta_2=0$. When $\Delta\omega_s=0$, which represents the case
when there is no cavity detuning, fidelity is 1.  As the two
cavities move apart, the spectra of the two cavities no longer
overlap. Thus, there is a small probability of photon detection at
$\hat{\textbf{d}}_{2}$ when the dipoles are in the state
$\ket{mm}$. This results in a loss of fidelity. Surprisingly,
however, the fidelity does not continue to decrease, but instead
increases back to 1 at some value of $\Delta \omega_s$.

As we keep increasing $\Delta\omega_s$ further, for a certain
value of the laser frequency $\omega$, both r$_1^g$ and r$_2^g$
are 0. If a detection event is observed in detector
$\hat{\textbf{d$_2$}}$, then the state of the two QDs collapses to
\begin{equation}
\begin{split}
&\ket{\Psi_f}_{dipoles}= \frac{1}{N}[(\alpha r_1^m-\beta
r_2^m)\ket{mm}+\alpha r_1^m\ket{mg} -\beta r_2^m\ket{gm}]
\end{split}
\end{equation}\
where $N^2=|\alpha r_1^m-\beta r_2^m|^2+|\alpha r_1^m|^2+|\beta
r_2^m|^2$. In this special case there is a second matching
condition, given by  $\alpha r_1^m = \beta r_2^m$, that again
projects the two dipoles onto
$\ket{\Psi_-}=(\ket{gm}-\ket{mg})/\sqrt{2}$. It is this second
matching condition that results in the fidelity of 1 at the second
peak.  Our optimization algorithm naturally detects these two
optimal regions, and gives us the best performance in the
intermediate regime.  Thus, given any set of operating conditions
we have the ability to determine the best set of amplitudes and
input frequencies.  We note that in many cases fidelities
exceeding 0.95 can be achieved even with an 60 GHz detuning, which
is more than half a cavity linewidth. The fabrication of cavities
with resonance frequencies that are repeatable within a linewidth
is well within current technological capabilities.

\begin{figure}[!h]
\centering
\includegraphics[width = 75mm,height = 40mm]{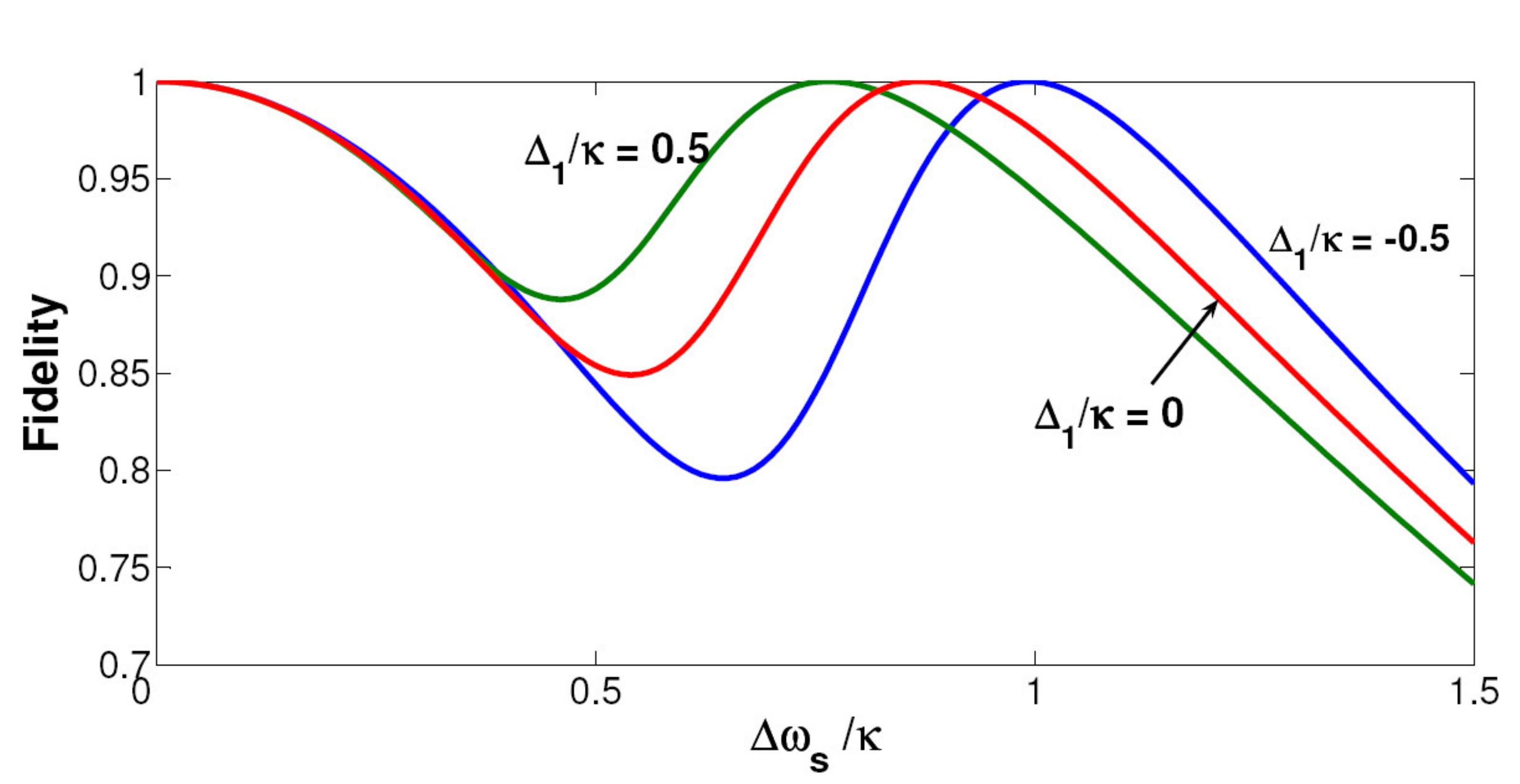}
\caption{a) Optimized fidelity as a function of $\Delta\omega_s$
for different values of $\Delta_1$. $\Delta_2 = 0$ }\label{ FIG 6}
\end{figure}

\begin{figure}[h]
\centering
\includegraphics[width = 75mm,height = 40mm]{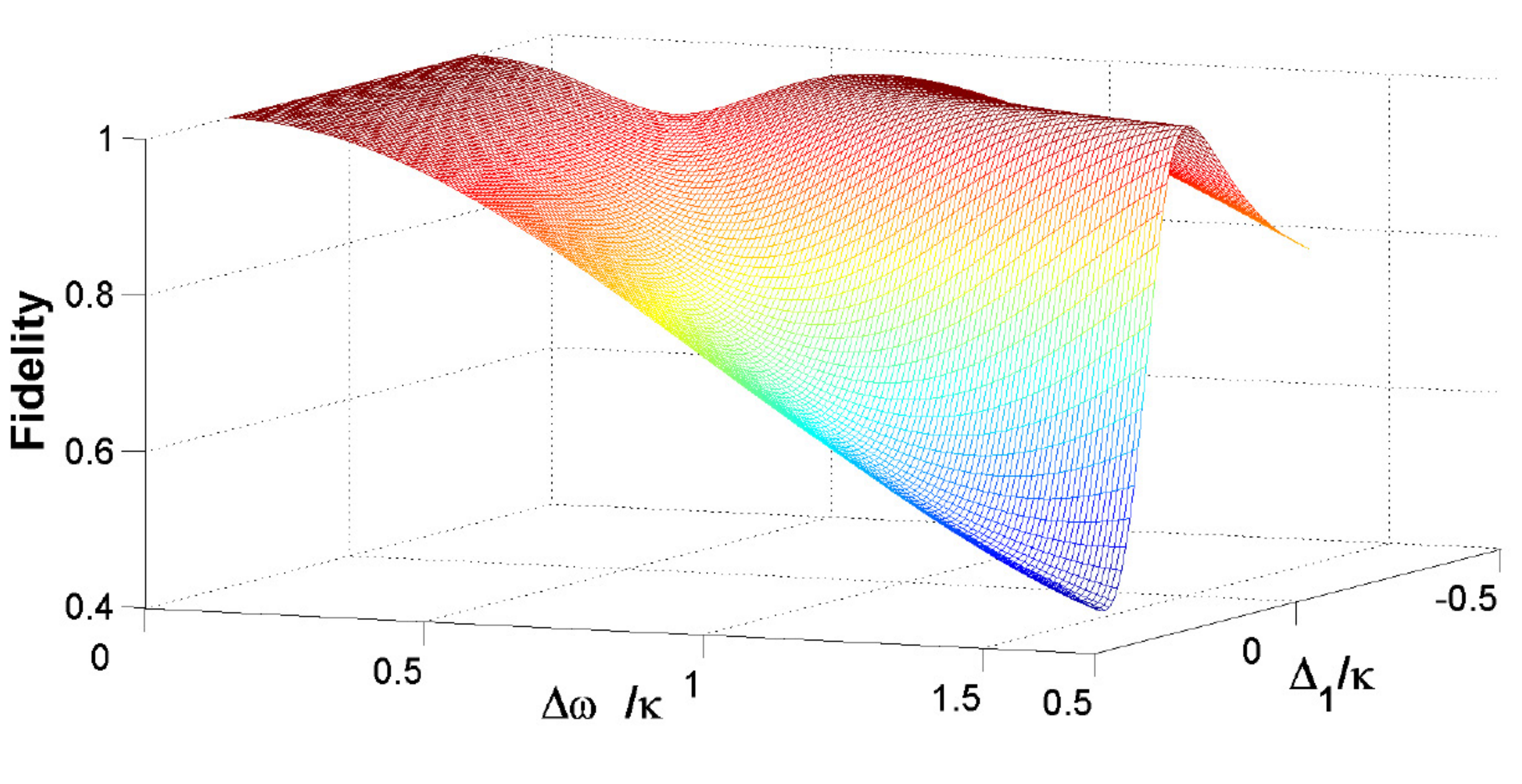}
\caption{(a) Fidelity as a function of cavity separations
$\Delta\omega_s$ and dipole detuning $\Delta_1$ }\label{FIG 6}
\end{figure}

  We can also consider what happens when we have both cavity detuning and dipole detuning.
  In Fig. 6, we plot optimized fidelity as a function of cavity
detuning $\Delta\omega_s$, and dipole detuning $\Delta_1$. A
maximum fidelity of 1 is obtained  when $\Delta\omega_s = 0$,
which represents the case when the two cavities have the same
resonant frequency. For $\Delta\omega_s<g$ (where g=20GHz in the
plot) the fidelity is largely independent of the detuning
$\Delta_1$ and is only determined by cavity separation. When
$\Delta\omega_s$ becomes larger, fidelity increases again due to
the second matching condition and a dependence on the dipole
detuning now becomes apparent.  This dependence on detuning comes
about from the fact that the second matching condition is a
function of $\Delta_1$, as illustrated in Fig. 5. From Fig 6,
fidelity is over 0.75 for a cavity linewidth separation(100 GHz)
of the cavities even over a wide range of dipole detunings. Thus,
we can use the protocol to obtain high fidelities even if the
cavities and dipoles are detuned.

\section{6. Exciton-Biexciton Implementation}

The protocol to generate entanglement between two QDs relies on
considering the QD as a three level system. One of the ways to
implement this three level system is by making use of the
excitonic and biexcitonic transitions.  A QD consists of three
states: the ground state, an exciton state X consisting of a
single electron-hole pair within the QD and a biexciton state XX
which is formed when two electron-hole pairs are trapped inside
the QD. The recombination of an electron-hole pair in the XX state
generates the biexciton XX photon. Similarly, the recombination of
an electron-hole pair in X state generates the X photon. The X and
XX photons have different energies due to the coulomb and exchange
interactions between the carriers. The typical energy separation
between the two lines is 1mev\cite{Gonz:2007}. Thus, we can make
use of this difference in energies to spectrally isolate the two
lines.

\begin{figure}[h]
\centering
\includegraphics[width = 75mm,height = 40mm]{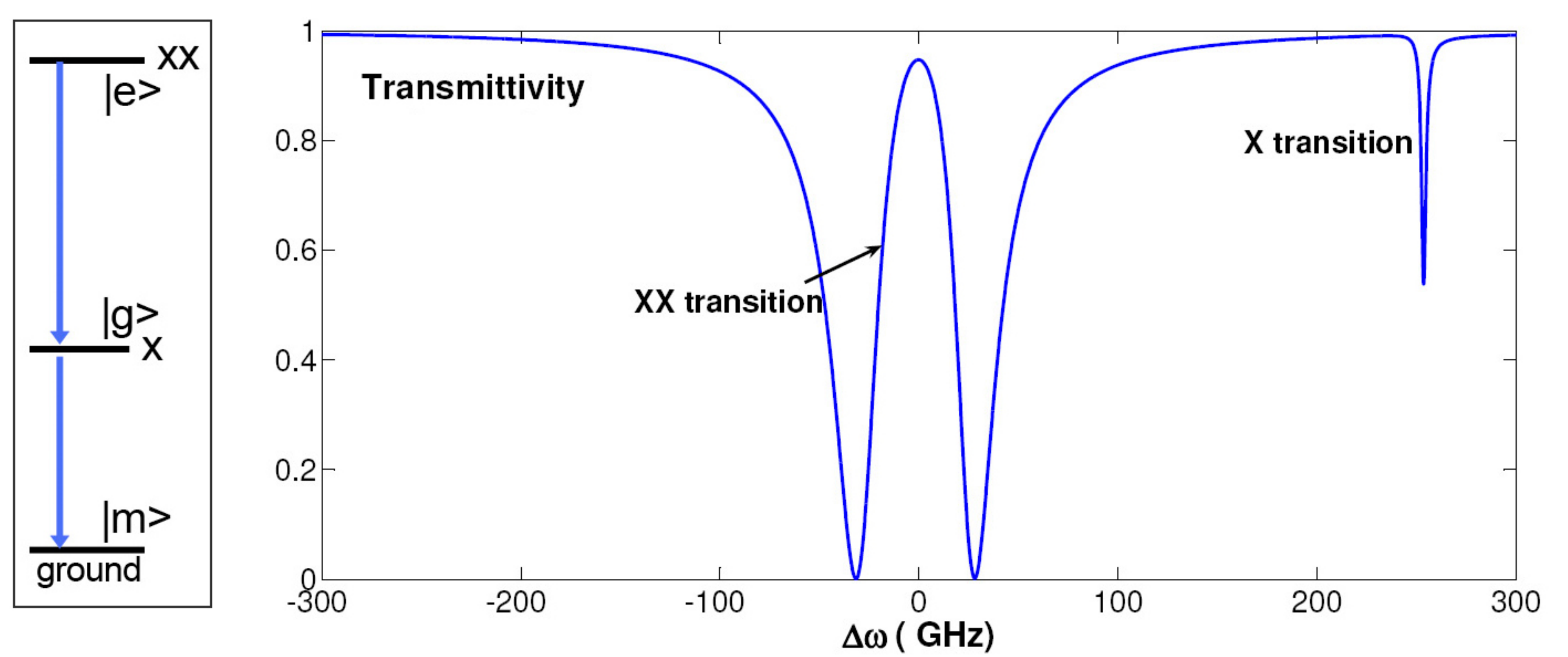}
\caption{ QD as a three level system }\label{fig7}
\end{figure}

The schematic of the QD as a three level system is shown in Fig
\ref{fig7}. We identify the three states of the QD as the ground,
X and XX states. We are free to assign these three states as
$\ket{g}$, $\ket{m}$ and $\ket{e}$ in a variety of different
combinations. In fact, there are several ways to assign these
levels, but probably the most convenient approach is given in the
inset of Fig \ref{fig7}. In the figure we have identified the
ground state of the QD as state $\ket{m}$, the single exciton
state as state $\ket{g}$, and the bi-exciton state as state
$\ket{e}$. This choice of the level configuration has a number of
advantages. First, single qubit operations between $\ket{g}$ and
$\ket{m}$ can be directly applied by pulses resonant with the
single exciton transition. Second, by placing the bi-exciton
transition on resonance with the cavity, we can enhance the
exciton to bi-exciton transition to get DIT, while at the same
time suppressing the single exciton lifetime in order to increase
the coherence time of the qubit.  This is illustrated in Fig
\ref{fig7}.

We assume the  biexciton transition to be on resonance with the
cavity frequency. This is indicated in Fig  \ref{fig7} where the
XX transition is in the middle of the cavity spectrum. The X
transition line is detuned from the cavity by $\delta_X$. The
vacuum Rabi frequencies of X and XX transitions are given by g$_X$
and g$_{XX}$ respectively. Similarly, the decay rates of the two
transitions is given by $\gamma_X$ and $\gamma_{XX}$ respectively.

In the exciton bi-exciton scheme the degree of cavity enhancement
directly impacts our ability to create an entangled state.  This
is because both the exciton and bi-exciton are strongly radiative
states, and the only way to enhance one while suppressing the
other is to use cavity lifetime modification.  In other qubit
implementations, such as dark state excitons\cite{Stevenson:2006},
this is not as much of a problem because selection rules make the
dark exciton long lived regardless of cavity.  To quantitatively
address this issue, we first calculate the coherence time of the
exciton state which is given by solving for the decay rate of
$\sigma_-$ in Eq. 1. The coherence time of the qubit is given by
\begin{equation}
\begin{split}
&\Gamma_{X} = \frac{g_X^2\kappa}{\delta_X^2+\kappa^2}+\gamma_X + \frac{1}{T_2}\\
\end{split}
\end{equation}
 where we
have added the dipole dephasing rate $1/T_2$ to the decay rate.
From the above equation one can see that increasing $\delta_X$
decreases the decoherence rate until it finally saturates at a
minimum value of $\gamma_X+1/T_2$.  At this point, increasing the
detuning of the exciton will not help as we are limited by
non-radiative and dephasing processes.

The coherence time of the dipole should be compared to an
appropriate time scale in order to determine if entanglement can
be generated.  Although there are a number of different factors
that should be considered in this comparison, the minimum
requirement for generating entanglement is that the duration of
the entangling pulse should be shorter than the coherence time of
the qubit.  If this is not the case, the qubit will begin to
decohere before the entangling pulse has finished interacting with
the cavity-dipole system, and there is no hope of generating
high-fidelity entanglement.  In previous work in Ref
\cite{Waks:2006}, it has been shown that when the pulse is
resonant with the dipole, it must be much longer than the modified
spontaneous emission lifetime of the dipole in order to be
monochromatic.  Thus, in the worst case when the dipole is
resonant with the cavity we need $1/\tau_p <<g_{XX}^2/\kappa$.
 We thus argue that an important figure
of merit is the ratio of the coherence time of the qubit to the
entanglement pulse width, given by
  \begin{equation}
    N_{ent} = \frac{g_{XX}^2}{\kappa\Gamma_X}
  \end{equation}
 This ratio determines the maximum
number of entanglements that can be performed before the system
decoheres. If $N_{ent}>1$, there is enough time for the pulses to
finish their interaction with the QDs before the system has
decohered.  Otherwise, the QDs will start to decohere before the
pulses have finished their interaction and high fidelity
entanglement will be impossible.

For calculations, we choose experimental values taken from the
paper of Hennessy et. al. \cite{Hennessy:2007} which investigates
the coupling of an Indium Arsenide (InAs) quantum dot coupled to a
photonic crystal cavity patterned in Gallium Arsenide (GaAs) by
electron beam lithography.  This experimental work reports g=20
GHz and cavity linewidth of 25 GHz which corresponds to a Q of
13300. However, the cavity linewidth is the bare cavity Q which
corresponds to the decay into the leaky modes. In order to achieve
critical coupling with the cavity, we need another in-plane mode
with a decay rate equal to the bare cavity decay rate. This mode
can be implemented in a photonic crystal as a waveguide coupled to
the cavity. Thus, the total decay rate of the cavity is double
that of the bare cavity decay rate.  Hence, we use $\kappa=50$ GHz
in our calculations. We use $g_X = g_{XX} = g$. For values of
T$_2$ we use 2 ns, which are appropriate values for InAs
QDs\cite{Langbein:2004}. For these values, $g_{XX}^2/\kappa$ is 8
GHz , $\Gamma_{x}$ is 0.93 GHz and $N_{ent}$ is 8.6. The fact that
$N_{ent}>1$ ensures that we can complete an entanglement operation
well before the QDs have decohered.

For a cavity linewidth of $50$ GHz, the exciton line lies outside
the cavity spectrum($\delta_{X} = 250$GHz). However, the exciton
line still couples to the cavity and we cannot ignore the presence
of the extra transition coupled to the cavity. So, we cannot
substitute for g as 0 in Eq 2 in order to obtain the cavity
reflection and transmission equations when the QD is in state
$\ket{m}$. We need to use the vacuum Rabi frequency as the value
for g to obtain the values of r$^m_1$, t$^m_1$, r$^m_2$ and
t$^m_2$. The changes in the transmission and reflection
coefficients will modify the final state of the QDs and hence the
fidelity of the system.

\begin{figure}[h]
\centering
\includegraphics[width = 75mm,height = 40mm]{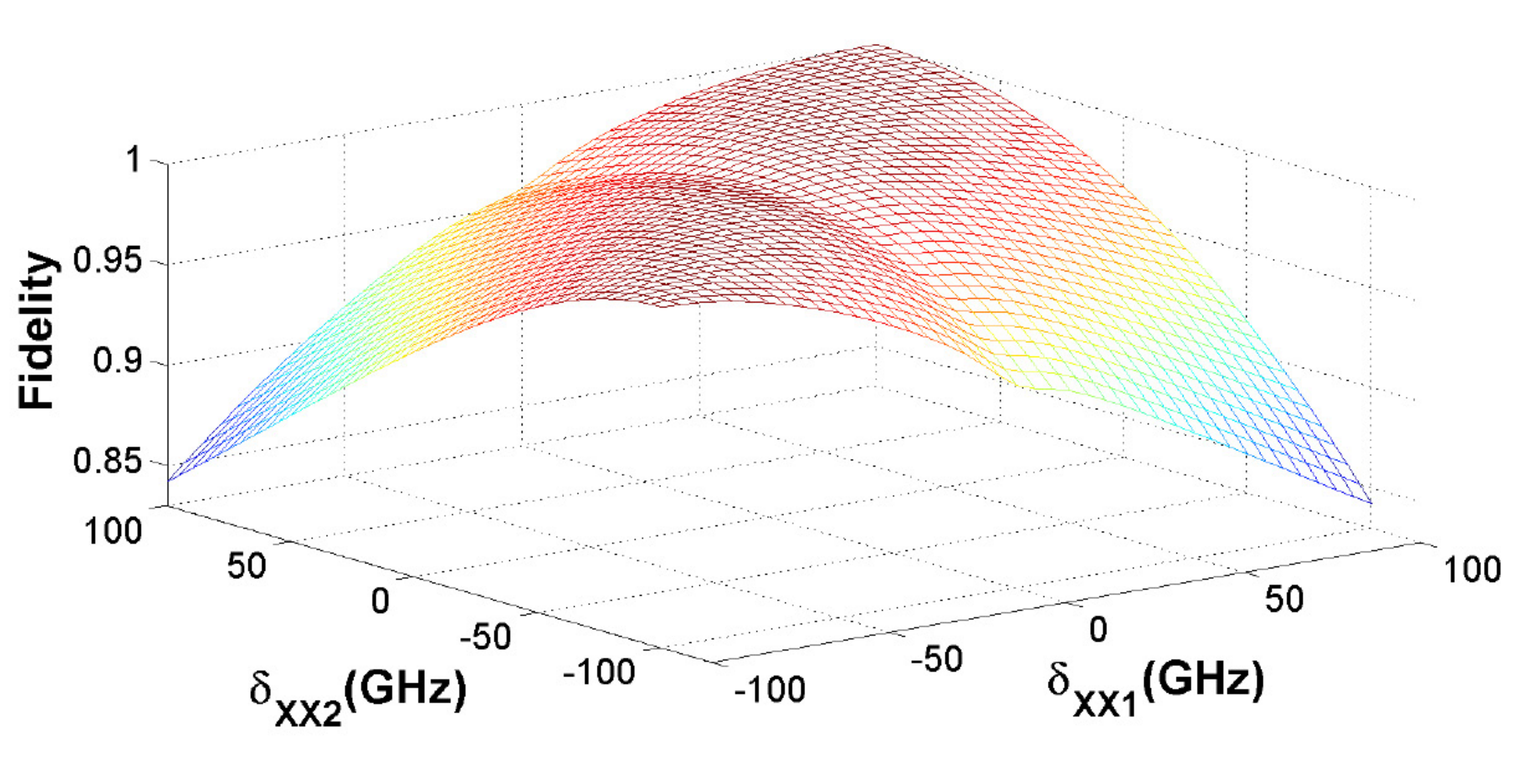}
\caption{ Fidelity as a function of dipole detunings $\delta_1$
and $\delta_2$ for the exciton-biexciton model of a QD
}\label{fig8}
\end{figure}

In general we cannot assume that the XX transition is not detuned
from the cavity spectrum. In order to see how robust the
biexciton-exciton protocol is dipole detunings, we define the
detunings of the XX transition lines from their cavities as
$\delta_{XX1}$ and $\delta_{XX2}$.  In Fig \ref{fig8} we plot the
dependence of fidelity on dipole detunings $\delta_{XX1}$ and
$\delta_{XX2}$ for the above case. For both $\delta_{XX1}=0$ and
$\delta_{XX2}=0$, fidelity is 1 as expected. When we increase
$\delta_{XX1}$ and $\delta_{XX2}$, the transmission and reflection
coefficients are modified due to the coupling of the X transition
to the cavity. This lowers the fidelity of the output state.  The
drop is fidelity is gradual and for a cavity linewidth separation
of the dipoles from the cavity resonance( 50 GHz), fidelity drops
to only 0.96. As we further increase the detunings to 100 GHz,
fidelity drops to 0.85. Thus, even for large detunings between the
cavities and the dipoles, reasonable high fidelity(0.85) states of
the QDs can be obtained. Thus, the exciton-biexciton scheme can be
used to create entanglement between QDs even if the exciton line
couples to the cavity. The performance of the protocol can be
further improved by fabricating cavities with high quality
factors.

\section{7. Conclusions}

 In conclusion, we have shown that one can achieve high fidelity entangled states between two dipoles,
  even when their emission frequencies are different.
  The method is robust to dipole and cavity frequency mismatch. Efficiency
  loss for a cavity linewidth change in dipole detuning is about
  50$\%$ for a constant fidelity. Therefore, relatively high
  fidelity can be obtained over a large range of dipole detunings
  without significant loss of efficiency.
     The development of protocols that are robust to these imperfections is extremely important for semiconductor
      based implementations of quantum networks.

      The authors acknowledge the support  of Army Research Office under
      the Grant No. W911NF0710427.


\section{APPENDIX}

The input coherent fields are $\ket{\alpha}$ and
      $\ket{\beta}$. Both the dipoles are initialized in a
      superposition of states $\ket{g}$ and $\ket{m}$. Thus, the
      initial state of the system is

\begin{equation}
      \ket{\Psi_i} = \ket{\alpha}\ket{\beta}(\ket{g} + \ket{m})(\ket{g} + \ket{m})/2
      \end{equation}

      The coherent states can be replaced by their corresponding
      displacement operators to account for all order of photon
      numbers. Thus,

      \begin{equation}
      \begin{split}
      \ket{\Psi_i} = &\frac{1}{2}e^{(\alpha \hat{\textbf{a}}^{\dagger}_{in} - \alpha^{*}
      \hat{\textbf{a}}_{in})}\ket{0}_{a_{in}}
       e^{(\beta \hat{\textbf{c}}^{\dagger}_{in} - \beta^{*} \hat{\textbf{c}}_{in})}\ket{0}_{c_{in}}\\
       &(\ket{gg}
       +\ket{gm} + \ket{mg}
+\ket{mm})\\
\end{split}
      \end{equation}

      The input fields after interactions with the cavity-dipole
      systems are transformed according to Eq 2.
      Thus, when the dipoles are in state $\ket{gg}$

 \begin{equation}
      \begin{split}
      &\hat{\textbf{a}}^{\dagger}_{in} \rightarrow r_1^g
      \hat{\textbf{a}}^{\dagger}_{out}+ t_1^g
      \hat{\textbf{b}}^{\dagger}_{out}\\
      &\hat{\textbf{c}}^{\dagger}_{in} \rightarrow r_2^g
      \hat{\textbf{c}}^{\dagger}_{out}+ t_2^g
      \hat{\textbf{d}}^{\dagger}_{out}\\
      \end{split}
      \end{equation}

Similar transformation equations apply when the dipoles are in the
states $\ket{gm}$, $\ket{mg}$ and $\ket{mm}$. The reflected field
from the two cavities is mixed on a 50/50 beamsplitter that
applies the transformation:

\begin{equation}
\begin{split}
&\hat{\textbf{a}}^{\dagger}_{out}\rightarrow(\hat{\textbf{d}}_{1}+\hat{\textbf{d}}_{2})/\sqrt{2}\\
&\hat{\textbf{c}}^{\dagger}_{out}\rightarrow(\hat{\textbf{d}}_{1}-\hat{\textbf{d}}_{2})/\sqrt{2}\\
\end{split}
\end{equation}

Applying the cavity and beamsplitter transformations on the
initial state $\ket{\Psi_i}$ we get

\begin{equation}
\begin{split}
\ket{\Psi_{{gg}}} = &D(\frac{\alpha r_1^g + \beta
r_2^g}{\sqrt{2}})D(\frac{\alpha r_1^g - \beta
r_2^g}{\sqrt{2}})D(\alpha t_1^g)D(\beta
t_2^g)\\
&\ket{0}_{d_1,d_2,b_{out},d_{out}}\ket{gg}\\
\end{split}
\end{equation}

$\ket{\Psi_{{gg}}}$ is the state of the output modes for the
dipoles in state $\ket{gg}$. This state can be split up into the
modes of detector $\hat{\textbf{d$_1$}}$, $\hat{\textbf{b}}_{out}$
and $\hat{\textbf{d}}_{out}$ and detector $\hat{\textbf{d$_2$}}$.
Thus,

\begin{equation}
\label{dipstate}
\begin{split}
&\ket{\Psi_{{gg}}}=\ket{\psi_{{gg}}}\ket{\mu_{gg}}\ket{gg}\\
&\ket{\psi_{{gg}}}= D(\frac{\alpha r_1^g + \beta r_2^g}{\sqrt{2}})D(\alpha t_1^g)D(\beta t_2^g)\ket{0}_{d_1,b_{out},d_{out}}\\
&\ket{\mu_{gg}}=D(\frac{\alpha r_1^g - \beta r_2^g}{\sqrt{2}})\ket{0}_{d_2}\\
\end{split}
\end{equation}

$\ket{\psi_{{gg}}}$ is the state of the output modes at detector
$\hat{\textbf{d$_1$}}$ and the transmitted modes
$\hat{\textbf{b}}_{out}$  and $\hat{\textbf{d}}_{out}$ when the
dipoles are in state $\ket{gg}$. $\ket{\mu_{gg}}$ is the field
amplitude at detector $\hat{\textbf{d$_2$}}$ when the dipoles are
in state $\ket{gg}$. Similarly, we can obtain the field amplitudes
$\ket{\Psi_{{gm}}}$, $\ket{\Psi_{{mg}}}$ and $\ket{\Psi_{{mm}}}$
when the dipoles are in states $\ket{gm}$, $\ket{mg}$ and
$\ket{mm}$.  The final state of the system is given by

\begin{equation}
\ket{\Psi_{{f}}}=\ket{\Psi_{{gg}}}+\ket{\Psi_{{gm}}}+\ket{\Psi_{{mg}}}+\ket{\Psi_{{mm}}}
\end{equation}

These states  $\ket{\Psi_{{gm}}}$, $\ket{\Psi_{{mg}}}$ and
$\ket{\Psi_{{mm}}}$ can be further decomposed on similar lines to
Eq \ref{dipstate} to obtain the field amplitudes
$\ket{\psi_{{gm}}}$ and $\ket{\mu_{gm}}$,$\ket{\psi_{{mg}}}$ and
$\ket{\mu_{mg}}$ and $\ket{\psi_{{mm}}}$ and $\ket{\mu_{mm}}$
respectively.

We define the projection matrix M as
$\sum_{n=1}^\infty\ket{n}_{d_2}\bra{n}$. M can also be written as

\begin{equation}
\begin{split}
M &= \sum_{n=0}^\infty\ket{n}_{d_2}\bra{n} - \ket{0}_{d_2}\bra{0}\\
    &= I - \ket{0}_{d_2}\bra{0}\\
\end{split}
\end{equation}

\begin{equation}
\rho_{dipoles}=\frac{tr_{\mathrm{(fields)}}\{\braket{M}{\Psi_f}\braket{\Psi_f}{M}\}}{tr_{\mathrm{(dipoles
\& fields)}}\{\braket{M}{\Psi_f}\braket{\Psi_f}{M}\}}
\end{equation}

\begin{equation}
\begin{split}
F & =  \bra{\Psi_-}{\rho_{dipoles}}\ket{\Psi_-}\\
   &=  \bra{\Psi_-}{\frac{tr_{\mathrm{(fields)}}\{\braket{M}{\Psi_f}\braket{\Psi_f}{M}\}}{tr_{\mathrm{(dipoles
\& fields)}}\{\braket{M}{\Psi_f}\braket{\Psi_f}{M}\}}}\ket{\Psi_-}\\
\end{split}
\end{equation}

The denominator is the probability of getting a detection at
detector $\hat{\textbf{d$_2$}}$. We identify this as efficiency
$\eta$.

\begin{equation}
\begin{split}
F & =  \bra{\Psi_-}{\rho_{dipoles}}\ket{\Psi_-}\\
   &= \bra{\Psi_-}{\frac{tr_{\mathrm{(fields)}}\{\ket{\Psi_f}\bra{\Psi_f}\}-tr_{\mathrm{(fields)}}\{\braket{0}{\Psi_f}\braket{\Psi_f}{0}_{d_2}\}}{\eta}}\ket{\Psi_-}\\
   &=\frac{F_1-F_2}{\eta}\\
\end{split}
\end{equation}

The individual terms can be evaluated to give

\begin{equation}
\begin{split}
&F_1 = \frac{1}{4}-\frac{\braket{\Psi_{{gm}}}{\Psi_{{mg}}}}{2} - \frac{\braket{\Psi_{{mg}}}{\Psi_{{gm}}}}{2}\\
F_2 =& \frac{1}{8}e^{-|\mu_{gm}|^2}+\frac{1}{8}e^{-|\mu_{mg}|^2}-\frac{1}{2}e^{-(|\mu_{gm}|^2+|\mu_{mg}|^2)/2}\braket{\psi_{mg}}{\psi_{gm}}\\
&-\frac{1}{2}e^{-(|\mu_{gm}|^2+|\mu_{mg}|^2)/2}\braket{\psi_{gm}}{\psi_{mg}}\\
&\eta = \frac{1}{4}[e^{-\mu_{gg}^2}+e^{-\mu_{gm}^2}+e^{-\mu_{mg}^2}+e^{-\mu_{mm}^2}]\\
\end{split}
\end{equation}

Thus, the complete expression for fidelity and efficiency can be
obtained.

\end{document}